\documentclass[aps,prl,reprint]{revtex4-1}

\usepackage{graphicx}  
\usepackage{dcolumn}   
\usepackage{bm}        
\usepackage{amssymb,amsmath}   

\usepackage[colorlinks=true,pdfstartview=Fit]{hyperref}		

\begin{document}


\title{Numerical modeling of inhomogeneous DNA replication kinetics}
\author{Michel G. Gauthier, Antoine Dub\'{e}, and John Bechhoefer}
\affiliation{Dept. of Physics, Simon Fraser University, Burnaby, BC,  V5A 1S6, Canada}
\date{\today}

\begin{abstract}
We present a calculation technique for modeling inhomogeneous DNA replication kinetics, where replication factors such as initiation rates or fork speeds can change with both position and time.  We can use our model to simulate data sets obtained by molecular combing, a widely used experimental technique for probing replication.  We can also infer information about the replication program by fitting our model to experimental data sets and also test the efficacy of planned experiments by fitting our model to simulated data sets.  We consider asynchronous data sets and illustrate how a lack of synchrony affects replication profiles.  In addition to combing data, our technique is well-adapted to microarray-based studies of replication. 
\end{abstract}

\maketitle

\section{Introduction}
\label{sec:intro}

The development of epifluorescence microscopy techniques in combination with single-molecule methods such as molecular combing has had an enormous impact on the study of DNA replication [see Refs.~\cite{Humana_chapter_Herrick, Herrick:2008p883, Herrick:1999p1088} for reviews].  Earlier approaches such as autoradiography~\cite{Cairns:1963p208} and 2D gel electrophoresis~\cite{Bell:1983p527} were not only slow but also provided only limited and localized information about the replication process.  The advent of reliable, high-throughput methods such as molecular combing led to genome-wide studies of DNA replication kinetics.  Moreover, the possibility of genome-level observations made possible complex experiments such as a study of the checkpoint regulation of DNA replication~\cite{Marheineke:2004p1086} and one of the relation between replication fork movement and origin spacing~\cite{Conti:2007p1424}. 

Molecular-combing measurements provide information about the DNA replication kinetics at a given time during the S phase~\cite{Bensimon:1994p1633}.  These snapshots of the replication process can be used to improve our knowledge about the location of replication origins, along with their activation time~\cite{Marheineke:2004p1086,Conti:2007p1424,Norio:2001p1594,Norio:2004p1566,Norio:2005p1567,Courbet:2008p1429,Katsuno:2009p1565, Pasero:2002p1639,Patel:2008p1640,Patel:2006p1634,Anglana:2003p1641, Herrick:2000p1396, Hyrien:2003p193,Herrick:2002p200}.  Such an analysis of the origin-firing program can, at least qualitatively, be based on a simple interpretation of combing measurements.  For example, even when studying cell cycles that are completely asynchronous, we can  infer from a series of combing measurements the relative replication times of different regions of the molecule.  This is of biological importance when considering both local and global changes in replication patterns that occur during normal cell differentiation, oncogenesis and oncogene-induced senescence~\cite{DiMicco:2007p154,DiMicco:2006p198},  Pluripotent embryonic stem cells, for example, rely on smaller replicon sizes that allow for diversification into distinct somatic cell lineages~\cite{Gondor:2009p1643}.  Knowledge of how replication patterns change during differentiation can give insight about underlying regulatory mechanisms and have contributed to our understanding of the impact of chromatin structure on technologies such as cloning~\cite{Lemaitre:2005p1644}.  Experiments based on synchronized cell populations can shed even more light on the origin positions and firing times, providing a detailed description of the space-time initiation program of DNA replication.         

In order to fully exploit the large data sets obtained from molecular-combing experiments, we must rely on computer-modeling techniques.  An approach commonly used to interpret combing results is to perform Monte Carlo simulations of hypothetical replication scenarios~\cite{Blow:2009p1454,Goldar:2008p982,Lygeros:2008p978, deMoura:2010p1630,Spiesser:2009p1631}.  In such cases, the models are not designed to reproduce all the biological details of DNA replication but rather are focused on the kinetic aspects of the process (i.e., the rate at which origins are activated or the speed at which DNA is duplicated by replication forks).  Nevertheless, fundamental biological features relating to the effect of chromatin modulations on replication patterns can be readily investigated.  Therefore, these simulations allow one to test hypotheses about the overall replication scenario and to quantify the kinetic parameters of a given scenario.  However, such simulations can be computationally intensive and therefore too slow to explore a series of scenarios (e.g., when performing a fit).  The numerical modeling approach we present here addresses this issue by allowing us to study the average replication kinetics without having to do simulations.  More precisely, our calculation technique gives the average replication rates and forks densities one would measure if an infinite number of replication cycles were observed or simulated.  The speed gain from using our method compared to simulations makes it possible to explore a variety of kinetic parameters and to fit experimental data.  This will facilitate automation of sample analysis that is of potential clinical and diagnostic interest because many genetic diseases such as cancer involve altered cell cycle and DNA replication kinetics~\cite{Kaufmann:2010p1604}.  

In the following sections, we will present our calculation technique and show how it can be used to study a given replication scenario.  Our modeling approach is flexible enough to accommodate inhomogeneous DNA replication kinetics, which are  scenarios in which the initiation rate and replication fork speed can change with position (and time) along the molecule.  Indeed, the methods presented here were motivated by single-molecule data that show clearly that the initiation of DNA replication in mammalian cells can occur in zones and that these zones are under developmental control~\cite{Norio:2005p1567}.  In addition, large-scale spatial variation of initiation rates and developmental control of such rates has been seen in many microarray-based experiments that give information about local, population-averaged replication rates [for example, \cite{Hiratani:2010p1642}].  Different chromatin environments, for example, correspond to regions with significantly different replication kinetics. The type of analysis discussed here can facilitate experiments addressing questions as to which chromatin factors (acetylation, methylation, phosphorylation) are involved in determining, on a concept specific basis, fork rates and origin efficiencies over extensive genomic regions.  

In the Materials Section, we will describe the type of combing data considered in the examples presented in this article.  The Methods Section will present the details of our calculation, as well as a comparison with simulated data.  This section will also show how our model can be adapted to study asynchronously replicating cells and how we can infer information about the replication program by fitting simulated combing data.  Finally, in the Notes Section, we will briefly discuss how a global analysis of all the replication kinetic parameters could be performed.  We will also discuss how the effects of fork blocks (or pause sites) can be included into our model.  This is also of potential clinical interest when examining the effects of a drug on DNA replication kinetics.  Recently, for example, it has been shown that sister replisomes decouple when they encounter damaged DNA~\cite{Doksani:2009p1515}.  Therefore, the number of unidirectional forks might be an indicator of the number of blocked replication forks in a normal cell cycle.  Such forks can be used as a marker to quantify and monitor transient DNA damage and DNA repair efficiencies on a genome-wide basis~\cite{Conti:2007p1475}.  This is just one example among many potential applications of this type of modeling.

\section{Materials and Methods}
\label{sec:materials}

\subsection{Kinetic description of DNA replication}
\begin{widetext}

\begin{figure}
\begin{center}
\includegraphics[width=12.0cm]{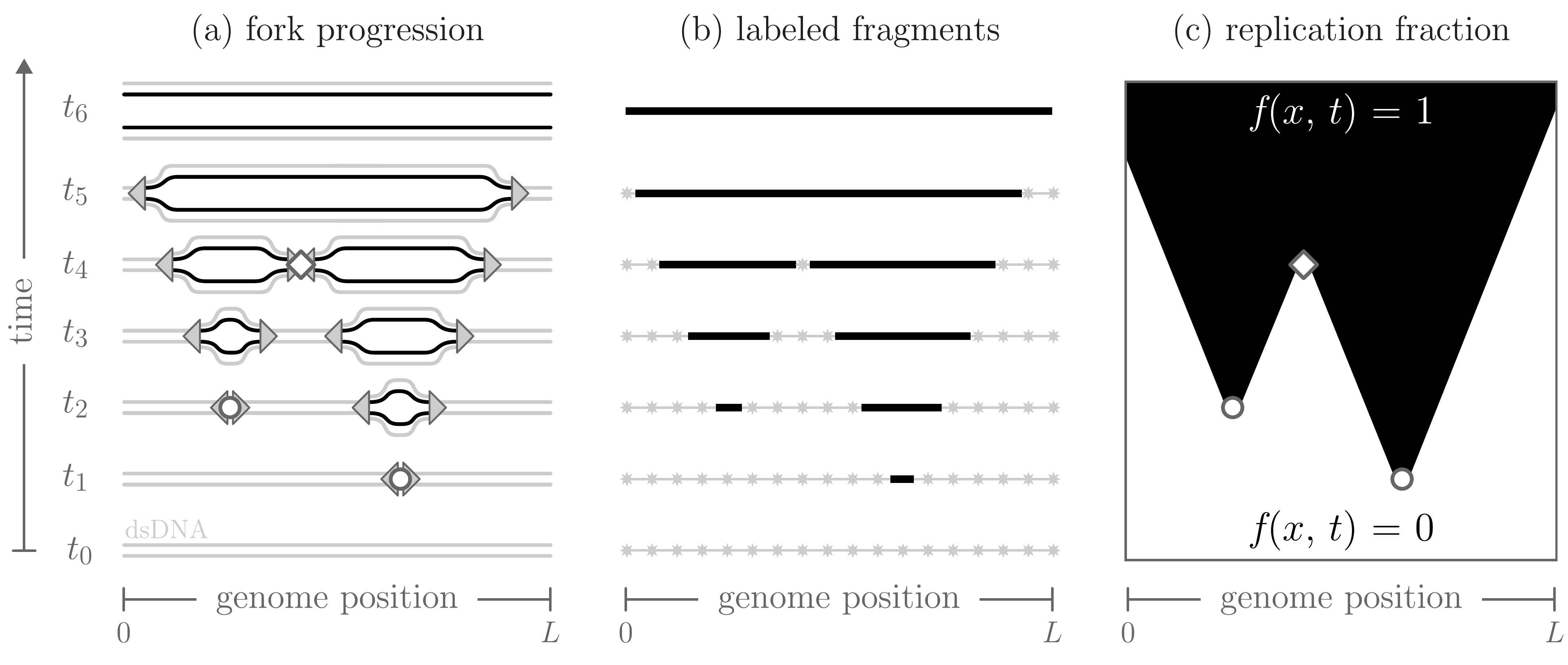}
\caption{(a) Schematic of the space-time progression of replication forks (triangles).  S phase starts at $t_0$.  Circles at times $t_1$ and $t_2$ show origins of replication, while the diamond indicates the termination site where forks from two origins coalesce at time $t_4$.  Replication forks move at  constant speed until they coalesce or reach the ends of the molecule of length $L$ (around $t_5$ and $t_6$).  Gray solid lines represent the original DNA molecule, while newly synthesized DNA is shown in black.  (b) Examples of labeled molecules collected during a combing experiment studying the replication kinetics presented in (a).  Gray stars indicate fluorescently labeled regions, while thick black lines correspond to label-free sections of the molecules.  Each molecule represents the labeling pattern one would observe if the nucleotides used for the DNA synthesis were replaced by fluorescent nucleotides at time $t_i$.  (c) Space-time replication fraction $f(x, t)$ of the replication cycle presented in (a).  White and black areas represent unreplicated and replicated DNA, respectively.         }  
\label{fig:schema}     
\end{center}
\end{figure}

\end{widetext}

The experimental situation we consider in this article is sketched in Fig.~\ref{fig:schema}.  Part (a) is a space-time illustration of the DNA replication kinetics during one cell cycle.  In this representation, replication can be initiated at various positions $x$ along the genome and at different times $t$ since the beginning of S phase (see the circles in Fig.~\ref{fig:schema}(a)).  A pair of oppositely moving replication forks emerges from each initiation site, and each fork starts to duplicate DNA at a speed $v(x, t)$. The dependence on $x$ and $t$ means that $v$ can change with both the location along the genome and the time during S phase.  For simplicity, the fork velocity used in this article is assumed constant throughout both space and time, as illustrated in Fig.~\ref{fig:schema}.  The forks progress until they reach one end of the molecule or until they coalesce with an oppositely moving fork (diamond symbol in Fig.~\ref{fig:schema}(a)).     

\subsection{Molecular labeling}
The labeling procedure consists in supplementing the regular nucleotides used for the DNA synthesis with nucleotide analogs
that can be detected by immunofluorescence microscopy.  If the cell cycles are synchronized, the nucleotide analogs can be introduced at a precise time $t_i$ after the beginning of S phase.  Figure~\ref{fig:schema}(b) shows examples of fluorescently labeled molecules that might be collected given the replication cycle presented in Fig.~\ref{fig:schema}(a).  Each example, from bottom to top, illustrates the labeled molecule obtained if the analogs were added at times $t_0$, $t_1$, $t_2$, $t_3$, $t_4$, $t_5$ or $t_6$.  Basically, this procedure labels the DNA that is replicated after the specific time $t_i$.  In an actual experiment, a second type of nucleotide analog would be used to map the rest of the molecule and hence obtain a fully labeled molecule.  In the simulations presented in this article, a second label is unnecessary, since there is no ambiguity about the molecule alignment or stretching.  The results we will present here focus on the \textit{replication fraction} of a molecule, $f(x,t)$, which is the probability that the section of the molecule located at the position $x$ along the genome is replicated by a given time $t$ during S phase. 

\subsection{Data collection}
\label{ss:data_collection}
In order to calculate the replication fraction $f(x, t)$, we must ensure that the experimentally collected labeled molecules meet the following conditions:
\begin{itemize}
\item the precise time when nucleotides were added must be known;
\item the molecules must be combed and stretched to identify the location of the labeled regions;
\item the molecules must be oriented (which can be achieved using hybridization probes).
\end{itemize}
Once these requirements are fulfilled, as in Fig.~\ref{fig:schema}(b), we can compute the replication fraction profile $f(x)$ of each labeled molecule by assigning the value $f=1$ (replicated at time $t$) to the non-labeled regions of the molecule and $f=0$ (unreplicated at $t$) to the labeled ones.  Figure~\ref{fig:schema}(c) presents the complete space-time replication profile resulting from our labeling procedure for the replication cycle example shown in Fig.~\ref{fig:schema}(a).  The labeling sequence can also be used to locate the positions and directions of replication forks as a function of time.  Although it is possible to analyze fork data, too, we focus here on the replication fraction [see Ref.~\cite{gauthier_submitted} for more on analyzing fork-density data].

In the following sections, we will compare calculated replication fractions to simulated ones.  Our simulated data sets mimic the data one could experimentally collect, in the sense that we simulate each step of a similar combing experiment.  On the other hand, our simulated data sets are ``better" in some ways than their experimental counterparts:
\begin{itemize}
\item the simulated cell cycles are perfectly synchronized;
\item the time when nucleotide analogs are introduced is known exactly (and once added, labels are instantly used for sequencing);
\item the positions along the simulated molecules are clearly defined (simulated results represent molecules that are perfectly stretched, aligned, and oriented).
\end{itemize}
Such ideal simulated data allow us to obtain precise statistics about the kinetics of replication itself, independent of experimental noise.  (We could also simulate the noise processes, but that would complicate the presentation here.)  As we will see below, these simulations can be used to test our numerical calculations.

\subsection{Monte Carlo simulations of combing experiments}
\label{ss:mcsim}
As mentioned above, we need to generate simulated data in order to test our numerical model of DNA replication.  The simulations we carried out follow the requirements listed in Section~\ref{ss:data_collection}.  For each simulated labeled molecule used in this article, we performed a Monte Carlo simulation of the replication dynamics as presented in Fig.~\ref{fig:schema}(a).  The numerical-simulation procedure we followed is presented in detail in Ref.~\cite{Jun:2005p149} and can be summarized as follows:
\begin{enumerate}
\item Define a simulation time step $\delta t$.  
\item Begin a cell-cycle simulation with an unreplicated molecule of size $L$ at time $t=0$, which marks the beginning of S phase.  
\item Simulate the initiation of replication origins (circles in Fig.~\ref{fig:schema}(a)) between $t$ and $t+\delta t$ using a pre-defined initiation rate $I(x, t)$.  This rate is defined as the number of initiations per time per length of unreplicated DNA.  
\item Allocate to each origin that fires a pair of oppositely moving replication forks (triangles in Fig.~\ref{fig:schema}(a)).  
\item Calculate the displacement of all existing replication forks during the time step $\delta t$ using the fork velocity profile $v(x, t)$.  Fork movement corresponds to DNA replication.        
\item Delete forks that coalesce with oppositely moving forks (diamond in Fig.~\ref{fig:schema}(a)) or that reach the end of the molecule.
\item Repeat from Step 3 until the replication of the molecule is completed. 
\item Define the time when the fluorescent nucleotides were introduced and deduce the corresponding labeled molecule pattern (as illustrated in Fig.~\ref{fig:schema}(b) for various label addition times).  As discussed below, this step is the one when we distinguish between synchronously and asynchronously replicating cells.    
\item Repeat from Step 2 to collect more molecules. 
\end{enumerate}

The simulation steps described above are valid if all cell cycles are perfectly synchronous.  For asynchronously replicating cells, we need to estimate the distribution of S phase start times (for example, using a Gaussian probability function).  Then, cell cycles themselves can be simulated following Steps 1--7 above.  However, the time when fluorescent nucleotides are added must be corrected relative to the S-phase start-time distribution.  In other words, we must keep track of the \textit{laboratory clock} compared to the \textit{cell-cycle clock}.

\subsection{Rate-equation approach}
\label{ss:req}
Our approach for calculating replication kinetics is based on a set of three coupled rate equations that describe the average space-time evolution of replication variables such as the replication fraction, $f(x, t)$, or the replication fork density, $\rho(x, t)$ (defined as the average number of forks per length at location $x$ and time $t$).  As we will see, the equations can be numerically solved in order to obtain the mean-field replication kinetics of a given replication scenario.  

The first equation of our set describes the rate of change of the probability that a given position along the genome is replicated as a function of time, 
\begin{equation}
	\frac{\partial f}{\partial t} = (v_{-} \,\rho_{-}) + (v_{+} \,\rho_{+}) \,,
	\label{e:dfdt}
\end{equation}
where the $\pm$ subscripts indicate the fork propagation direction (for right- and left-moving forks, respectively). Note that Table~\ref{tab:1} presents  the definition and units of all symbols used in this article.  Equation~\ref{e:dfdt} simply states that the rate of change of the replication fraction is given by the product of the local fork densities times the rate at which each fork replicates.  Note that the velocity of left- and right-moving forks can be different in our formalism, although here they are kept equal.      

The second and third equations of our set express the space-time rate of change of the right- and left-moving fork densities as 
\begin{equation}
	\label{e:drhodt_r}
	\frac{\partial\rho_{+}}{\partial t}  + \frac{\partial (v_{+} \rho_{+})}{\partial x} = I (1-f) - \frac{(v_{-}+v_{+}) \rho_{-} \rho_{+}}{1-f}  \,,
\end{equation}
and 
\begin{equation}
	\label{e:drhodt_l}
	\frac{\partial\rho_{-}}{\partial t}  - \frac{\partial (v_{-} \rho_{-})}{\partial x} = I (1-f) - \frac{(v_{-}+v_{+}) \rho_{-} \rho_{+}}{1-f}  \,,
\end{equation}
respectively.  The left-hand sides of these equations describe the change of fork densities due to their displacement along the genome.  The right-hand sides present the contributions associated with initiations and coalescences of forks.  The first term is simply given by the product of the local initiation rate, $I(x, t)$, normalized by the probability that the genome is not already replicated at position $x$ .  The second term gives the frequency at which converging replication forks meet at $x$ and $t$.  This coalescence rate is proportional to the local densities of both types of forks and their relative speed when they merge.  This rate must also be normalized by the probability that the local DNA is not already replicated, $1-f(x, t)$.  For clarity, we did not explicitly specify in Eqs.~\ref{e:dfdt}--\ref{e:drhodt_l} that $f(x, t)$, $\rho_\pm(x, t)$, $v_\pm(x, t)$ and $I(x, t)$ can all change with both the position along the genome, $x$,  and the time elapsed from the beginning of S phase, $t$. 

Given a replication scenario for the initiation rate, $I(x, t)$, and the fork propagation speed, $v_\pm(x, t)$, Eqs.~\ref{e:dfdt}--\ref{e:drhodt_l} can be numerically solved for $f(x, t)$ and $\rho_\pm(x,t)$.  The initial conditions must be set to $f(x, 0)=0$ and $\rho_\pm(x,0)=0$ for all positions along the molecule.  Such a numerical analysis can be carried out using various approaches.  We developed and ran our own numerical code using IGOR~Pro~\cite{igor}.  In the code, we discretize the modeled space-time region and approximate the partial derivatives in the equations describing the replication kinetics by finite differences.  Knowing the initial conditions for all positions at time $t=0$, one can then numerically integrate the system as a function of time [details about such numerical methods can be found in Ref.~\cite{nr2007}, for example].  The results presented in this article were obtained using the single-step explicit Euler method.  We used $dx=0.002$~Mb and $dt=0.0 \bar{1}$~hr (note that we must have $dx/dt \geq v$ in order to solve the system).  For simulating or fitting to larger regions of the genome, the added efficiency of a more sophisticated numerical method for integrating the partial differential equations could be worth the extra effort to implement.

\subsection{Boundaries and finite-size fragments}
\label{ss:boundary}
When modeling a whole genome using our calculation technique, one must specify the boundary conditions.  The modeling we have described so far applies to linear DNA where replications forks vanish when they reach the ends of the molecule.  Our simulation algorithm and calculation method can easily be adapted for the modeling of circular DNA molecules by using \emph{Periodic Boundary Conditions} (PBCs).  This simply means that any fork that reaches one end of our modeled domain (i.e., $x=0$ or $x=L$) is re-inserted at the other end.  In terms of our rate-equation system, using PBCs means imposing the conditions that $\rho_\pm(0, t)=\rho_\pm(L, t)$ at all times $t$.  

Molecular-combing experiments are often designed to extract DNA fragments from a particular region of the genome [e.g. \cite{Norio:2005p1567,Lebofsky:2006p1627}].  In order to model such fragments, we have two options:
\begin{enumerate}
\item We can model the whole chromosome and extract the information about the region that represents the experimentally observed fragment.
\item We can model only the fragment of interest, adding boundary terms to include replication by forks that initiate outside the fragment and then enter it during S phase. 
\end{enumerate}
The first option is a straightforward application of the model presented in Section~\ref{ss:req} and has been applied, for example, to the study of Chromosome VI in yeast (\textit{S. cerevisiae})~\cite{Czajkowsky:2008p152}.  However, if the chromosome is much longer than the fragment of interest, as is usually the case, it will be experimentally and computationally inefficient to study and model the whole molecule.  In such a case, the second option can be an excellent approach, as long as we can model the effective flux of incoming forks at the modeled domain boundaries.  The simplest situation is that no forks from outside the domain cross the boundaries.  However, such a situation is realistic only when the ``outside origins," if they exist, are far from the fragment's boundaries.  More generally, we can include the contribution of origins near the fragment region in our model by replacing the periodic boundary conditions with time-dependent functions at either side that represent the flux of forks coming from outside the domain.  The fork densities at the fragment boundaries are then~\cite{gauthier_submitted}
\begin{eqnarray}
	\rho_+ &= \beta_{+}(t)\,[1-f_+]  \quad \text{(left edge)}
	\,, \nonumber \\[3pt] 	
	\rho_- &= \beta_{-}(t)\,[1-f_-]  \quad \text{(right edge)} \,,
\label{e:rho_boundaries}
\end{eqnarray}  
where $\rho_+ = \rho(x=0,t)$, $f_+ = f(x=0,t)$, $\rho_- = \rho(x=L,t)$, and $f_- = f(x=L,t)$.  The functions $\beta_\pm(t)$ characterize the injection of right- and left-moving forks at $x=0$ and $x=L$, respectively.  For example, if the initiation rate outside the left and right edges is constant ($I_\pm$, respectively) in the semi-infinite region beyond the analyzed fragment, and if the fork velocity is constant as well, then $\beta_\pm(t) = I_\pm \, t$~\cite{gauthier_submitted}.  However, the real advantage in focusing on the fork fluxes $\rho_\pm$ is that we do not need to have a detailed scenario of replication outside the region where data are collected but we only need to find a form for the time dependence of $\rho_\pm$ that fits the data well.  If it turns out that $\rho_\pm = I_\pm t(1-f_\pm)$ fits the data well, then we can assert that origin initiation outside the region, insofar as it affects the region under study, is equivalent to a zone of constant intitiation rate.  In practice, we have found that the functional form $\rho_\pm = I_\pm t(1-f_\pm)$, with $I_\pm$ as fit parameters, works well, because the closest origin or origin cluster to the boundary tends to dominate.

\subsection{Post-treatment of the rate-equation solution}
Once the solution of our rate-equation system is obtained for a given replication scenario [$I(x,t)$, $v(x,t)$, boundary conditions, ...], we can use it to calculate the space-time distributions of initiations and coalescences.  These probability profiles tell us where and when such events are more likely to occur, deepening our knowledge about the replication kinetics.  

The initiation density function is given by
\begin{equation}
\label{e:phi_i}
	\phi_i = \frac{I(1-f)}{N_i} \,,
\end{equation}
where 
\begin{equation}
\label{e:N_i}
	N_i=\int_0^\infty\int_0^LI(1-f)\,dx\,dt \,,
\end{equation}
is the average number of initiations per replication cycle between $x=0$ and $x=L$.  The initiation density function is thus simply a normalized version of the initiation term in Eqs.~\ref{e:drhodt_r} and~\ref{e:drhodt_l}.  Similarly, we can normalize the coalescence term to obtain the space-time probability density of observing a coalescence as
\begin{equation}
	\label{e:phi_c}
	\phi_c =\frac{(v_{-}+v_{+}) \rho_{-} \rho_{+}}{N_c(1-f)} 	\,,
\end{equation}
with the average number of coalescences per cycle given by 
\begin{equation}
	N_{c}=\int_0^\infty\int_0^L\frac{(v_{-}+v_{+}) \rho_{-} \rho_{+}}{(1-f)} \,dx\,dt \,.
\label{e:N_c}
\end{equation}  
Other useful replication-related information can be derived from our rate-equation solution.  For example, we can use $f(x, t)$ and $\rho_\pm(x, t)$ to derive the distributions of the first initiation or the last coalescence times [see Ref.~\cite{gauthier_submitted} for more details].

\section{Results and Discussion}
\subsection{Model validation}
\begin{figure}
\begin{center}
\includegraphics[width=9cm]{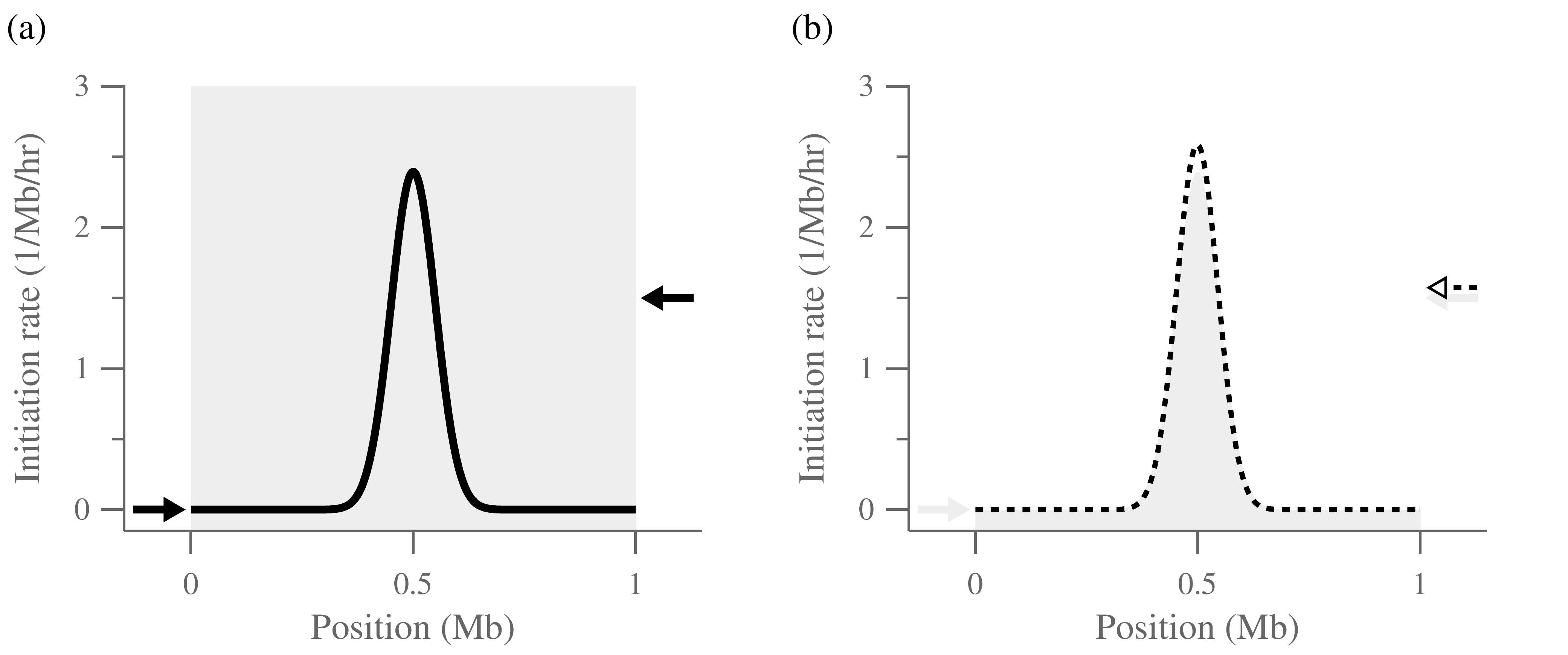}
\caption{(a) Initiation profile $I(x)$ used in the simulations and calculations presented here.  The solid line represents the number of initiations per unreplicated length of DNA per unit time as a function of the position along the genome.  The curve itself is defined as a $I(x)=I_0 e^{-(x-\mu)^2/2\sigma^2}/\sqrt{2\pi\sigma^2}$, with $I_0=0.3$~hr$^{-1}$, $\mu=0.5$~Mb and $\sigma=0.05$~Mb.  The arrows at $x=0$ and $x=L$ indicate the level of initiation outside the modeled region (shaded in light gray), $I_+=0$~Mb$^{-1}$hr$^{-1}$ and $I_-=1.5$~Mb$^{-1}$hr$^{-1}$ (see Section~\ref{ss:boundary}).  
(b) Fit (dashed line and arrow) of the eight simulated replication profiles for synchronous cells presented in Fig.~\ref{fig:sync_vs_async}(a).  The fit is compared to the original initiation scenario from part (a), which is shown in light gray.  The parameters obtained from the fit are $I_0=0.301$~hr$^{-1}$, $\mu=0.499$~Mb, $\sigma=0.0464$~Mb and $I_-=1.57$~Mb$^{-1}$hr$^{-1}$
}
\label{fig:init_rate}     
\end{center}
\end{figure}

We now compare our rate-equation solution to simulation results.  The origin-initiation scenario we chose is presented in Fig.~\ref{fig:init_rate}(a).  It consists of a Gaussian-shaped initiation zone at the center of a 1~Mb fragment.  We used the boundary conditions derived at the end of Section~\ref{ss:boundary}, i.e., a constant level of initiation outside the fragment region [parameters for $I(x,t)$ are presented in caption of Fig.~\ref{fig:init_rate}].  In order to simplify the interpretation of the results, we assumed that the initiation profile of Fig.~\ref{fig:init_rate}(a) does not change with time, although our modeling formalism allows such variation.  Similarly, we used a constant fork velocity ($v=0.15$~Mb/hr $= 2.5$~kb/min).  Again, our formalism also works equally well if the fork velocity depends on space and/or time.

\begin{figure}
\begin{center}
\includegraphics[width=8.0cm]{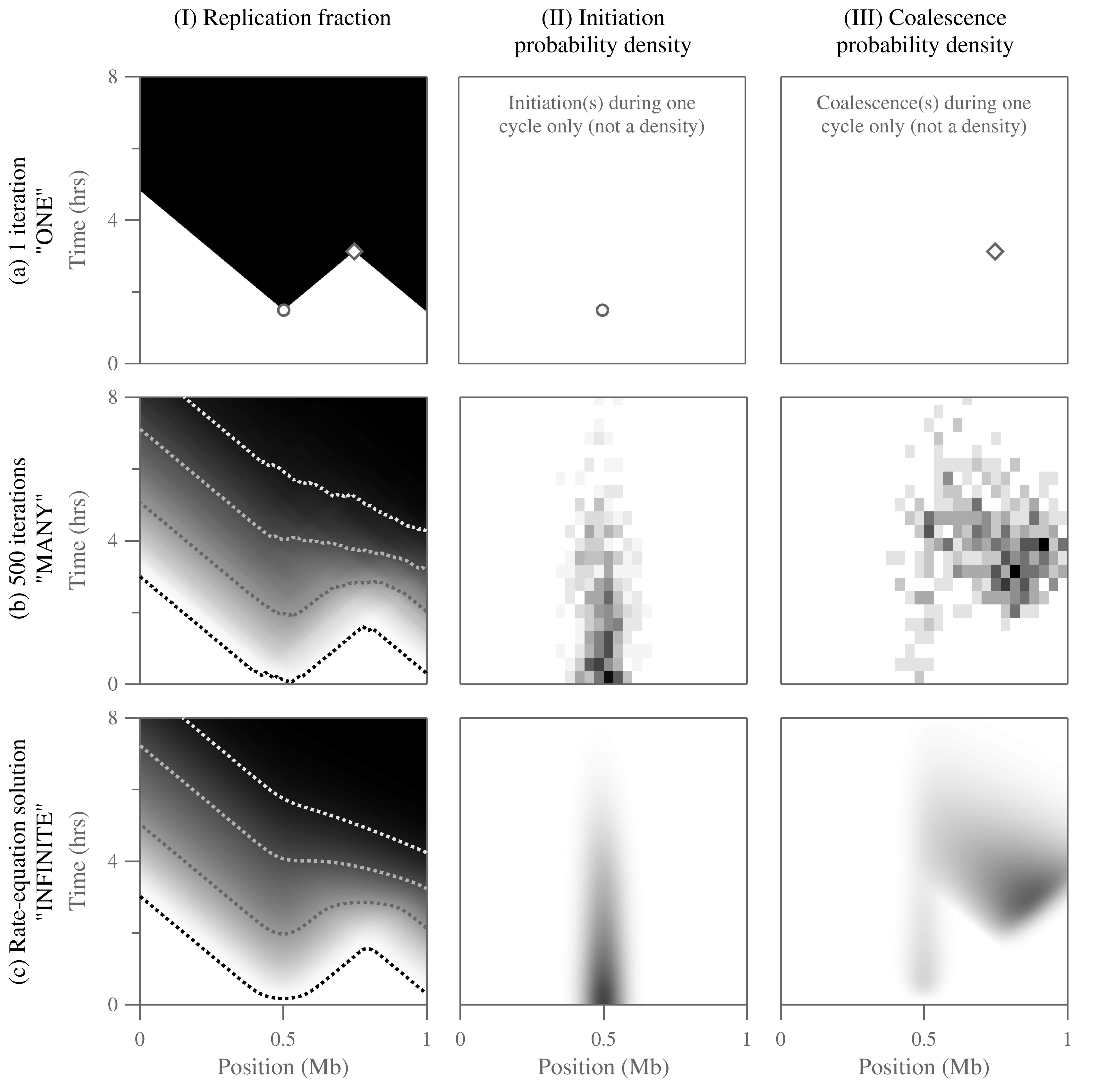}
\caption{Simulated replication cycles (a, b) compared to our rate-equation solution (c).  In part (I), the shading indicates the value of the replication fraction, $f(x, t)$, from 0 (white) to 1 (black).  The dashed lines in (b-I) and (c-I) represent 1\%, 40\%, 70\% and 90\% contour curves (from bottom to top).  Contour curves show the time when the genomic locations reach a given probability of being replicated.   Parts (II) and (III) indicate the likelihood of initiations and coalescences, respectively.  Part (a) shows individual initiations and coalescences from one cycle, while parts (b) and (c) give the normalized probability density functions of observing such events as a function of space and time, $\phi_i(x, t)$ and $\phi_{c}(x, t)$.  The gray scales go from 0 (white) to $3.6$/Mb/hr (black) for initiations and from 0 (white) to $1.8$/Mb/hr (black) for coalescences.  Both the simulation and the calculation were done using the initiation function $I(x)$ presented in Fig.~\ref{fig:init_rate}(a) (constant throughout time).  The fork velocity was set to $v(x, t)= 0.15$~Mb/hr.  
  }
\label{fig:sim2REQ}     
\end{center}
\end{figure}

Figure~\ref{fig:sim2REQ} compares the calculated replication fraction and the initiation and the coalescence probability densities with simulation data.  Part (I) of that figure helps to understand the physical meaning of our rate-equation solution.  The figure illustrates that the calculated replication profile corresponds to the average profile we would get were we able to perform an infinite number of simulation iterations (the mean-field solution).  Figure~\ref{fig:sim2REQ} shows that our calculated initiation and coalescence probability density functions agree with simulation results as well.  More specifically, we used Eqs.~\ref{e:N_i} and~\ref{e:N_c} to calculate the average numbers of initiations and coalescences observed within the modeled domain per replication cycle (we obtained $N_i=0.91$ and $N_c=0.84$).  These calculated values agree, within statistical fluctuations, with the simulation averages of $N_i=0.91\pm0.04$ and $N_c=0.82\pm0.04$, based on 500 simulated replication cycles.            

Finally, note that although the number of simulated replication cycles is large, the average profiles presented in part (b) are still relatively noisy.  Using rate equations gives us access to the exact average replication kinetics, whereas a simulation always gives a noisy approximation.  Reducing the noise to usable levels can take a large number of simulations.   On the other hand, simulations are good for understanding the level of statistical noise in a given experiment (by simulating the exact number of molecular fragments collected, etc.) and can also give information about more subtle features, such as correlations in the fluctuations of different measurements.

\subsection{Modeling a synchronous combing experiment}
\begin{figure}
\begin{center}
\includegraphics[width=8.0cm]{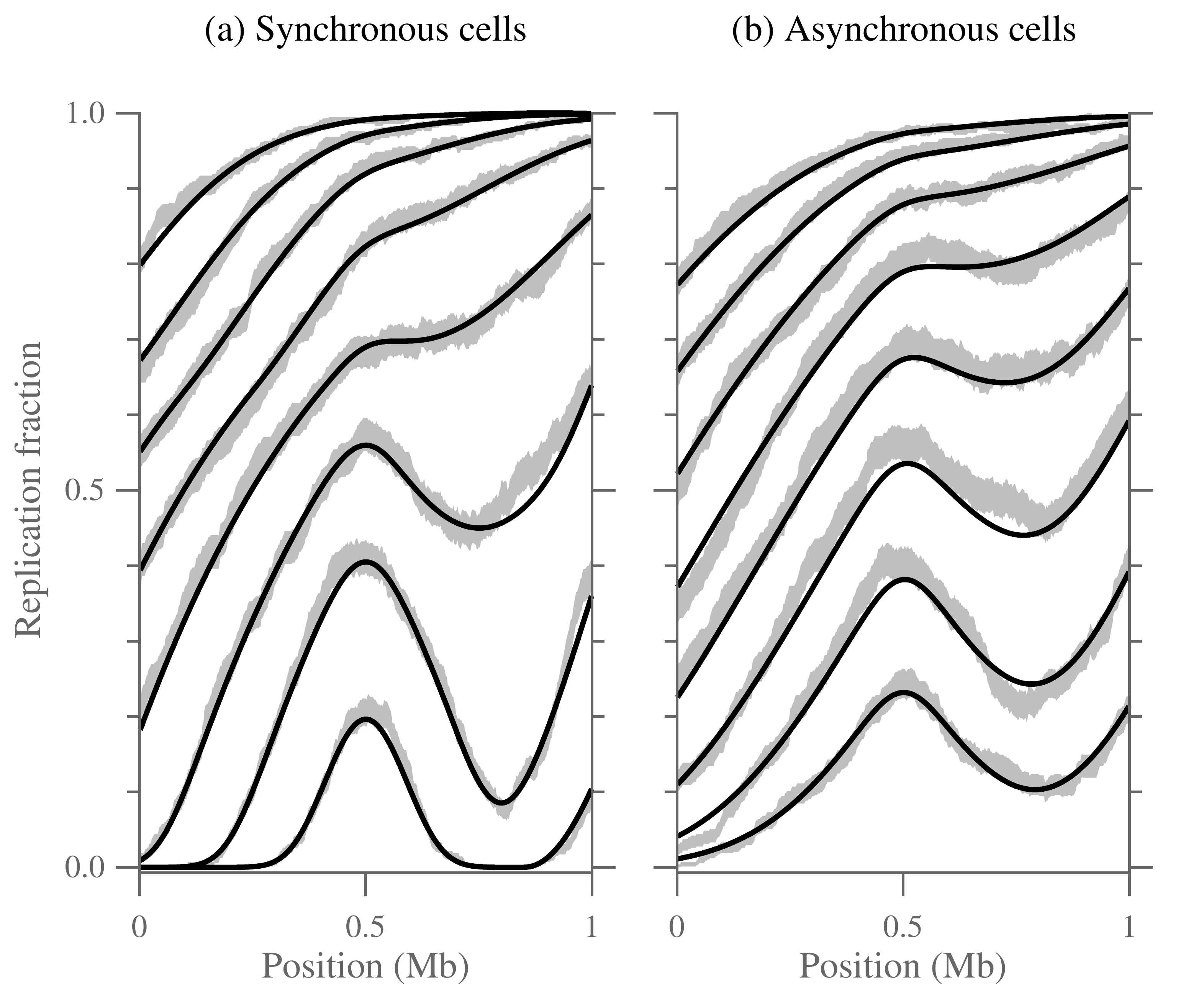}
\caption{(a) Average combing profile (or replication fraction) at a series of time points (from bottom to top: 1, 2, 3, 4, 5, 6, 7 and 8~hrs) for synchronous replication cycles.  The shaded areas represent the confidence intervals of the average simulated combing profiles (the average value $\pm$ the standard error of the mean, obtained from 500 iterations).  The solid lines show the rate-equation solution for $f(x)$ at the corresponding times (the same solution presented in Fig.~\ref{fig:sim2REQ}(c-I)).  (b) Similar data for asynchronous replication cycles.  The start-time distribution was the same as in Fig.~\ref{fig:sync2async}).  The simulated data were collected as in part (a) except that the start times of the cycles were randomly chosen.  The rate-equation solution was modified to account for the start-time asynchrony (see text and Fig.~\ref{fig:sync2async} for details).      }
\label{fig:sync_vs_async}       
\end{center}
\end{figure}

As described in Section~\ref{ss:mcsim}, our Monte Carlo algorithm can mimic a simple combing experiment of synchronously replicating cells.  The gray bands in Fig.~\ref{fig:sync_vs_async}(a) present simulated combing results obtained using the replication scenario described in Fig.~\ref{fig:init_rate}(a).  These bands represent the average simulated combing profiles for various incorporation times of the nucleotide analogs (from 1 to 8~hrs).  For each band, 500 combed molecules were simulated; the gray bands shown in Fig.~\ref{fig:sync_vs_async}(a) are the average replication profiles $\pm$ the standard error of the means.  

These simulation results can be directly compared to the calculated mean-field replication fraction, $f(x,t_{i})$, where $t_i$ is the analogs supplementing time.  The calculated combing profiles are presented as solid lines in Fig.~\ref{fig:sync_vs_async}(a) and agree with the simulated ones within statistical limits.  

\subsection{Correction for cell-cycle asynchrony}
In the previous section, we demonstrated that our rate-equation approach is consistent with averaged combing results when all cell cycles are perfectly synchronous.  We now show how our calculated replication fraction $f(x, t)$ can be modified in order to take into account the asynchrony of S phase start times in the population of measured cells.  We consider two cases: (1) the distribution of start times is known, and (2) cell cycles are completely asynchronous.

\subsubsection{S phase start-time distribution}
\begin{figure}
\begin{center}
\includegraphics[width=9cm]{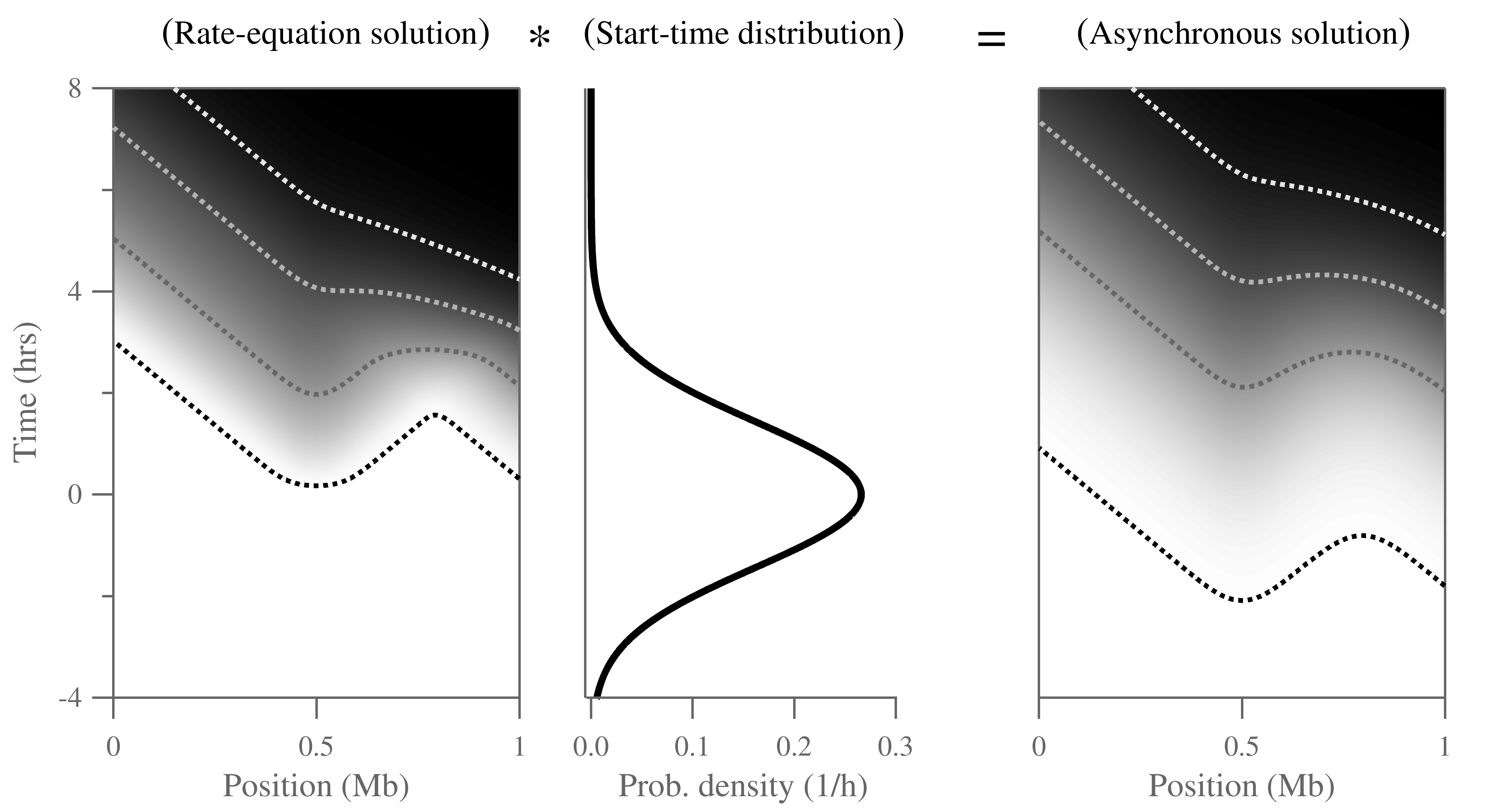}
\caption{The rate-equation solution for $f(x,t)$ can be convoluted with the replication start-time distribution to obtain the average replication profile of asynchronously replicating cells.  Left:  Representation of the same solution as shown in Fig.~\ref{fig:sim2REQ}(c-I) with an extended time axis.  Middle: Probability density function of the start time of S phase for asynchronized cells.  The distribution is Gaussian, with standard deviation $\sigma_{\mathrm{start}}=1.5$~hr).  Right:  Transformed replication fraction profile, $f_{\mathrm{async}}(x, t)$ (see Eq.~\ref{e:fasync}).  As in Fig.~\ref{fig:sim2REQ}, the dashed lines indicate the 1\%, 40\%, 70\% and 90\% contour curves.    }
\label{fig:sync2async}     
\end{center}
\end{figure}

Let $t$ refer to the laboratory clock and let $\phi_{\mathrm{start}}(t)$ be the S phase start-time probability density function.  If $\phi_{\mathrm{start}}(t)$ is known (or assumed), we can convolute $\phi_{\mathrm{start}}(t)$ with our rate-equation solution for $f(x, t)$ to find the asynchronous replication curve:
\begin{align}
	f_{\mathrm{async}}(x, t) &= \phi_{\mathrm{start}}(t) *f(x, t) \nonumber \\
	&\equiv  \int_{-\infty}^{\infty} \phi_{\mathrm{start}}(t^\prime) f(x, t-t^\prime)dt^\prime \,,
	\label{e:fasync}
\end{align}
where $f_{\mathrm{async}}(x, t)$ is the average observed replication function.  The convolution operation is illustrated in Fig.~\ref{fig:sync2async} for our replication scenario solution example of Fig.~\ref{fig:sim2REQ}(c-I) and a Gaussian distribution for the cell-cycle start times.  The situation depicted here could correspond to an experiment where cells are imperfectly synchronized by drug treatment [see Ref~\cite{Herrick:2002p200} for an example].  In such a case, Eq.~\ref{e:fasync} can be used to correct our calculation.

The impact of an asynchronous start-time distribution on combing measurements is presented in Fig.~\ref{fig:sync_vs_async}(b).  For this figure, the same simulation procedure as in part (a) was followed except that a Gaussian distribution with a standard deviation of 1.5~hr was used to adjust the cell-cycle clock with respect to the laboratory time.  Again, the gray bands represent the average and standard error of the mean obtained from 500 simulations, while the solid lines represent calculated profiles, $f_{\mathrm{async}}(x, t_i)$.

\subsubsection{Perfectly asynchronous cells}
Equation~\ref{e:fasync} allows us to study a great number of partially synchronous cases but must be modified when the spread of start times becomes wider than the typical replication time.  
An extreme case is the limit of perfect asynchrony, where molecular fragments are collected at any time during the cell cycle, with a uniform probability distribution. 
In this limit, $f_{\rm asynch}(x,t) $ is constant for all $x$ and $t$ and no useful information about the replication kinetics can be inferred.
Norio~\textit{et~al.} introduced another way to analyze data obtained for this case of perfect asynchrony~\cite{Norio:2001p1594}.  The method uses only information from fragments that show evidence of active replication (i.e., that have at least one replication fork).  The main difficulty is that the length of S phase---the time between the first initiation and the last coalescence---is different for each cell because initiations are stochastic and vary from cell to cell.  Although the analysis is more involved and beyond the scope of this paper, it is still possible to extract replication parameters (initiation rates, fork velocities, etc.).  See Ref.~\cite{gauthier_submitted} for details.

\subsection{Fitting data}
\label{ss:inference}
An analytical formalism such as that presented in this article makes fast predictions of statistical profiles from a given replication scenario.  Using this mathematical approach, we can infer replication parameters from experimental data more quickly and precisely than using simulation-based fitting methods.  

To illustrate these possibilities, we performed a simple fit of the synchronous simulation data presented in Fig.~\ref{fig:sync_vs_async}(a) using the initiation profile as a fitting variable.  More precisely, we fitted for the three parameters, \{$I_0$, $\mu$, $\sigma$\}, used to describe the Gaussian initiation peak of Fig.~\ref{fig:init_rate}(a) and for the outside contribution, $I_-$, as well (the left side contribution, $I_+$, was set to zero since a free fit gave a value consistent with zero).  The eight experimental curves of Fig.~\ref{fig:sync_vs_async}(a) were simultaneously fit using a standard routine~\cite{igor}.  The best-fit initiation profile is presented and compared to the original one in Fig.~\ref{fig:init_rate}(b).  Adding parameters, we can also infer the fork velocity, $v$, and the standard deviation of S phase start times, $\sigma_{\mathrm{start}}$ (for asynchronous data).  Note the need to account for correlations in the noise fluctuations to evaluate correctly the statistical errors of the fit parameters [see Section~\ref{ss:fork} and Ref.~\cite{gauthier_submitted} for details].

\subsection{Fork density profiles}
\label{ss:fork}
DNA replication patterns, which depend on chromatin context, origin efficiencies and replication fork rates, are highly inhomogeneous both within the genome and across different tissue types.  Knowledge of these patterns and how they evolve is crucial to understanding regulation of the cell cycle in both normal and diseased cells.  As we discussed above, molecular-combing measurements provide replication fractions, $f(x)$, at a given time $t$ during S phase, and we have shown here that such profiles can be reproduced using our calculation technique.  In addition to the replication state of the molecule, combing experiments also indicate the location of replication forks, and hence their density, at the time of nucleotide-analog incorporation.  This additional information about fork densities can also be compared with our rate-equation solution.  Thus, we could improve the inference technique presented in Section~\ref{ss:inference} by simultaneously fitting the replication fractions and the two fork density functions using a common set of parameters.  However, such ``global fits" are not trivial.  The correlations between fluctuations in different parts of the data---for example, how replication fractions correlate with fork densities---must be taken into account~\cite{gauthier_submitted}.

\subsection{Fork arrest and DNA damage}
Tumor cells are characterized by extensive karyotype alterations and aberrant DNA replication kinetics associated with deletions and amplifications.  The altered kinetics has been found to depend on amplicon structure~\cite{Conti:2007p1475, Griffiths:1991p1535}, and recently a Werner syndrome cell line was shown to exhibit altered rDNA patterns and elevated densities of unidirectional replication forks.  The relationship between genetic structure and replication kinetics is poorly understood, but repeat-rich sequences in heterochromatin and elsewhere in the genome have an important impact on replication kinetics and genome stability~\cite{Rodriguez:2002p30}.  Our rate-equation techniques can be easily adapted to study the impact of stochastic fork arrests on the replication kinetics.  Using the model presented in this article, we can model the effect of a change in speed of the replication forks as they move along the genome, via the velocity profile $v(x, t)$.  Such an approach could be used to model zones where forks slow (for example, the replication slow zones discussed in~\cite{Cha:2002p1432}).  The same approach can also be used to model a local or global change in fork velocity due to a checkpoint response.  On the other hand, phenomena such as fork blocks due to DNA damage~\cite{Vilenchik:2003p1523}, where the fork velocity is different in each cell, need to be modeled differently.  Our model can be adapted to such situations.  For example, we have considered cases where forks stall at randomly located DNA lesions and may eventually restart after repair~\cite{gauthier_submitted,Gauthier:2010p1614}.  The main modification needed is to introduce two new fork densities, $\rho_\pm^\prime(x, t)$, which represent the densities of left- and right-moving forks that have stalled.  The rate of change of these new densities is governed by the stall rate of regular forks and the repair frequency of stalled forks.  The replication kinetics would then be described using five differential equations instead of three.  Two extra terms must also be added to the right-hand sides of Eqs.~\ref{e:drhodt_r} and~\ref{e:drhodt_l}, to represent the stall and repair rates.

\section{Conclusions}
In this paper, we presented a simple and versatile simulation technique to model inhomogeneous DNA replication kinetics. We demonstrated that our modeling approach can be used to analyze or simulate molecular combing data as well as microarray-based results. We showed that our model can also be used to infer information about the replication kinetics from fitting typical experimental measurements.  

\section*{Acknowledgments}
This work was supported by a Discovery Grant from the Natural Sciences and Engineering Research Council of Canada and by the Human Frontier Science Program.  We thank Paolo Norio, whose experimental results and cheerful collaboration pushed us to develop many of the results presented here.


\clearpage

\begin{table}
\caption{List of symbols}
\label{tab:1}       
\begin{tabular}{p{2.2cm}p{8.3cm}p{0.3cm}p{2cm}}
\hline\hline\noalign{\smallskip}
\textbf{symbol} & \textbf{definition}  & & \textbf{units}  \\
\noalign{\smallskip}\hline\hline\noalign{\smallskip \smallskip}

$x$ & position along the genome \smallskip \smallskip  && Mb \\
\hline
$t$ & time elapsed since the beginning of S phase  \smallskip \smallskip && hr \\
\hline
$L$ & genomic length of the modeled region \smallskip \smallskip && Mb \\
\hline
$f(x, t)$ & replication fraction ($0\leq f \leq 1$)\smallskip \smallskip && --- \\ 
\hline
$\rho_\pm(x, t)$ & right- and left-moving replication fork densities  \smallskip \smallskip && 1/Mb\\ 
\hline
$v_\pm(x, t)$ & right- and left-moving replication speeds \smallskip \smallskip && Mb/hr \\
\hline
$I(x, t)$ & number of initiations per length of unreplicated DNA per time \smallskip \smallskip &&1/Mb/hr \\
\hline
$I_\pm$ & initiation rates outside the modeled region ($I_+$ is used to calculate the flux of right-moving forks coming from the $x<0$ region while $I_-$ is used for the left-moving forks initiated at $x>L$)  \smallskip \smallskip && 1/Mb/hr\\
\hline
$\phi_{\{i,\,c\}}(x, t)$ &  probability density of observing an initiation or a coalescence \smallskip \smallskip && 1/Mb/hr \\
\hline
$N_{\{i,\,c\}}$ &  average number of initiations or coalescences per replication cycle \smallskip \smallskip && --- \\
\hline
$\phi_{\mathrm{start}}$ &  S phase start-time probability density function of asynchronous cells \smallskip \smallskip && 1/hr \\
\hline
$\sigma_{\mathrm{start}}$ &  standard deviation of the S phase start times of asynchronous cells \smallskip \smallskip && hr \\
\hline
$f_{\mathrm{async}}(x, t)$ &  transformed replication fraction profile for asynchronous cells \smallskip && --- \\

\noalign{\smallskip}\hline\hline\noalign{\smallskip}
\end{tabular}

\end{table}

\end{document}